\documentclass[pdflatex,bst/sn-mathphys-num]{sn-jnl}

\usepackage{graphicx}
\usepackage{amsmath,amssymb,amsfonts}
\usepackage{amsthm}
\usepackage{mathrsfs}
\usepackage{xcolor}
\usepackage{booktabs}
\usepackage{url}
\usepackage{hyperref}

\usepackage{bm}
\usepackage{amstext}
\usepackage{braket}
\usepackage{esint}
\usepackage{colortbl}
\usepackage{float}
\usepackage{flafter}
\usepackage{placeins}
\usepackage{tikz}
\usepackage{tikz-cd}
\usepackage{quantikz}

\definecolor{qGreenFill}{HTML}{E7F4E4}
\definecolor{qGreenDraw}{HTML}{2E7D32}
\definecolor{qBlueFill}{HTML}{E7F0FF}
\definecolor{qBlueDraw}{HTML}{1E5AA8}
\definecolor{qMeasFill}{HTML}{F2F2F2}
\definecolor{qMeasDraw}{HTML}{666666}

\tikzset{
  qboxgreen/.style={draw=qGreenDraw, fill=qGreenFill, rounded corners=2pt, line width=0.4pt,
                    inner xsep=6pt, inner ysep=3pt, align=center},
  qboxblue/.style ={draw=qBlueDraw,  fill=qBlueFill,  rounded corners=2pt, line width=0.4pt,
                    inner xsep=6pt, inner ysep=3pt, align=center},
  meter/.append style={draw=qMeasDraw, fill=qMeasFill, rounded corners=1.5pt, line width=0.35pt}
}

\graphicspath{{./images/}{./figures/}}

\begin{document}
% =========================
% Title + Authors (sn-jnl)
% =========================
\title[Distilling ECGFounder into a VQC Student]{From Foundation ECG Models to NISQ Learners: Distilling ECGFounder into a VQC Student}

\author*[1]{\fnm{Giovanni} \sur{dos Santos Franco}}\email{gs.franco@unesp.br}
\author[1]{\fnm{Felipe} \sur{Mahlow}}\email{f.mahlow@unesp.br}
\author[2,3]{\fnm{Ellison Fernando} \sur{Cardoso}}\email{Ellison.Cardoso@einstein.br}
\author*[2,1]{\fnm{Felipe} \sur{Fanchini}}\email{felipe.fanchini@einstein.br}

\affil[1]{\orgdiv{Faculty of Sciences}, \orgname{São Paulo State University (Unesp)}, \orgaddress{\city{Bauru}, \country{Brazil}}}
\affil[2]{\orgname{Hospital Israelita Albert Einstein}, \orgaddress{\city{São Paulo}, \country{Brazil}}}
\affil[3]{\orgdiv{Laboratory of Magnetic Resonance in Neuroradiology (LIM-44)}, \orgname{University of São Paulo Faculty of Medicine Clinics Hospital}, \orgaddress{\city{São Paulo}, \country{Brazil}}}
\abstract{Foundation models have recently improved electrocardiogram (ECG) representation learning, but their deployment can be limited by computational cost and latency constraints. In this work, we fine-tune ECGFounder as a high-capacity teacher for binary ECG classification on PTB-XL and the MIT-BIH Arrhythmia Database, and investigate whether knowledge distillation can transfer its predictive behavior to compact students. We evaluate two classical 1D students (ResNet-1D and a lightweight CNN-1D) and a quantum-ready pipeline that combines a convolutional autoencoder, which compresses 256-sample ECG windows into a low-dimensional latent representation, with a 6-qubit variational quantum circuit implemented in Qiskit and executed in a simulated backend. Across both datasets, the teacher provides the strongest overall performance, while distillation yields competitive students under a considerable reduction in trainable parameters. We further analyze the sensitivity of student performance to distillation settings, highlighting consistent accuracy--efficiency trade-offs when compressing a foundation ECG model into classical and quantum-ready learners under a unified evaluation protocol.}

\maketitle

\section{Introduction}\label{sec:introduction}

Electrocardiography (ECG) remains a primary, non-invasive modality for assessing cardiac function, and the increasing availability of large-scale digital ECG archives has accelerated the development of data-driven methods for automated interpretation. \cite{Hannun2019Cardiologist}
Deep learning, particularly convolutional neural networks, has demonstrated strong performance for rhythm and arrhythmia recognition, supporting the feasibility of learning-based ECG decision systems in clinically relevant settings. \cite{Kim2025HybridCNNTransformerArrhythmia}
This progress has been further enabled by openly available benchmarks that permit transparent and reproducible evaluation across studies and cohorts.
Among these, PTB-XL provides a large, clinically annotated dataset, while the MIT-BIH Arrhythmia Database remains a long-established reference for arrhythmia detection and classification. \cite{Wagner2020PTBXL,Moody2001MITBIH}

More recently, representation learning via foundation-model pretraining has emerged as a principled strategy to leverage massive corpora and obtain transferable features that can be adapted to downstream tasks through fine-tuning. \cite{McKeen2025ECGFM}
ECGFounder operationalizes this paradigm by pretraining on over ten million ECG recordings with broad label coverage and by emphasizing external evaluation across multiple clinical settings. \cite{Li2024ECGFounder}
Despite consistent gains, the deployment of high-capacity models may be limited by memory footprint, computational cost, and latency requirements, particularly in resource-constrained environments. \cite{Hinton2015Distilling}
Knowledge distillation (KD) provides a practical route to address these constraints by transferring a teacher’s predictive behavior to smaller students, typically via softened targets that convey informative inter-class structure beyond one-hot supervision; distillation can also be strengthened through intermediate representation matching when student capacity is highly constrained. \cite{Hinton2015Distilling,Romero2015FitNets}
In parallel, KD has begun to be explored in quantum machine learning as a mechanism to transfer knowledge across heterogeneous model classes under explicit quantum resource constraints, including classical-to-quantum distillation using frozen classical teachers and strategies that compress quantum models while preserving their behavior. \cite{Hasan2023BridgingKD,Alam2022KDQNN}

In this study, we fine-tune ECGFounder as a high-capacity teacher for binary ECG classification on PTB-XL and the MIT-BIH Arrhythmia Database, and then distill the resulting teacher into compact students to quantify accuracy--efficiency trade-offs across learning paradigms. \cite{Li2024ECGFounder,Wagner2020PTBXL,Moody2001MITBIH,Hinton2015Distilling}
We consider two classical 1D students (a compact ResNet-style model and a lightweight CNN) and a quantum-ready student based on a convolutional autoencoder coupled to a shallow variational quantum circuit, enabling a controlled comparison under a unified evaluation protocol. \cite{He2016ResNet,Schuld2018CircuitCentric,Preskill2018NISQ}
The main contributions of this work are:

\begin{itemize}
\item To distill a state-of-the-art ECG foundation model (ECGFounder) into compact student models for binary ECG classification on PTB-XL and MIT-BIH, quantifying the resulting accuracy--efficiency trade-offs;
\item To assess the performance of a VQC-based student within this distillation framework, using a quantum-ready pipeline (convolutional autoencoder + 6-qubit VQC) under a controlled evaluation protocol.
\end{itemize}

The remainder of this paper is organized as follows. Section~\ref{sec:methodology} reviews the background on ECGFounder, knowledge distillation, and variational quantum circuits. Section~\ref{sec:method} details the proposed methodology, including preprocessing, teacher fine-tuning, student architectures, and the training and evaluation protocol. Section~\ref{sec:results} presents the quantitative results and analyzes the impact of distillation settings. Finally, Section~\ref{sec:conclusion} concludes the paper and outlines directions for future work.

% =========================
% ECGFounder (Model Overview)
% =========================
\section{Theoretical Background}\label{sec:methodology}

\subsection{ECGFounder}\label{sec:ecgfounder}

ECGFounder is a foundation model for electrocardiography (ECG) designed to learn transferable representations from large-scale ECG corpora and adapt them to downstream clinical tasks via fine-tuning. \cite{Li2024ECGFounder}
This direction is motivated by the strong empirical success of deep learning for arrhythmia detection and ECG classification, together with the growing availability of public ECG benchmarks that enable reproducible evaluation across cohorts and acquisition protocols. \cite{Hannun2019Cardiologist,Wagner2020PTBXL,Moody2001MITBIH}
At a high level, ECGFounder follows a representation-learning pipeline in which an ECG encoder maps an input segment to a compact embedding, and lightweight task-specific heads can be attached for supervised objectives such as classification. \cite{Li2024ECGFounder}
Pretraining is performed at scale with broad label coverage, encouraging the encoder to capture morphology, rhythm, and temporally meaningful features that can be reused across tasks, thereby reducing the need for extensive task-specific feature engineering in downstream settings. \cite{Li2024ECGFounder,Hannun2019Cardiologist}

A key emphasis of ECGFounder is external, multi-domain evaluation: rather than optimizing for a single benchmark, the framework stresses transfer across datasets and clinical settings to probe robustness under dataset shift—an important concern in ECG due to differences in patient populations, acquisition devices, and annotation conventions. \cite{Li2024ECGFounder,Wagner2020PTBXL,Moody2001MITBIH}
Because foundation models typically incur substantial deployment costs (memory, latency, and compute), they are commonly paired with compression methods—most notably knowledge distillation—so that smaller students can preserve teacher behavior while meeting real-world constraints. \cite{Hinton2015Distilling}
When desired, distillation can be further strengthened by supervising intermediate representations in addition to output targets, improving the transfer of internal feature structure under tight capacity budgets. \cite{Romero2015FitNets}

% =========================
% Knowledge Distillation
% =========================
\subsection{Knowledge Distillation}\label{sec:kd}

KD is a compression paradigm in which a compact \emph{student} is trained to approximate the predictive behavior of a high-capacity \emph{teacher}, using the teacher’s outputs as an additional supervision signal. \cite{Bucilua2006ModelCompression,Hinton2015Distilling,Gou2021KDSurvey}
The central idea is that, beyond the correct class, the teacher distribution encodes informative structure about class similarities, which the student can exploit to improve learning under limited capacity. \cite{Hinton2015Distilling,Gou2021KDSurvey}

Let $\mathbf{v}\in\mathbb{R}^{C}$ and $\mathbf{z}\in\mathbb{R}^{C}$ denote teacher and student logits for $C$ classes. KD typically introduces a temperature-scaled softmax to control the \emph{softness} of these targets:
\begin{equation}
p_i^{(T)}=\frac{\exp(v_i/T)}{\sum_{j=1}^{C}\exp(v_j/T)},\qquad
q_i^{(T)}=\frac{\exp(z_i/T)}{\sum_{j=1}^{C}\exp(z_j/T)},
\label{eq:kd_softmax_temp}
\end{equation}
where $T>1$ increases the entropy of the teacher distribution, amplifying the relative contribution of non-argmax classes and exposing information contained in logit magnitudes that is largely suppressed at $T=1$. \cite{Hinton2015Distilling,Ba2014DoDeepNetsNeed,Gou2021KDSurvey}
During training, the student is optimized to match the teacher distribution at the same temperature $T$; at inference time, predictions are computed with $T=1$. \cite{Hinton2015Distilling,Gou2021KDSurvey}

When ground-truth labels $\mathbf{y}$ are available, KD is commonly implemented by mixing a soft-target term with the standard supervised objective:
\begin{align}
\mathcal{L}_{\text{soft}}(T) &= -\sum_{i=1}^{C} p_i^{(T)} \log q_i^{(T)}, \label{eq:kd_soft_ce}\\
\mathcal{L}_{\text{hard}} &= -\sum_{i=1}^{C} y_i \log q_i^{(1)}, \label{eq:kd_hard_ce}\\
\mathcal{L}_{\text{KD}} &= (1-\alpha)\,T^{2}\,\mathcal{L}_{\text{soft}}(T) + \alpha\,\mathcal{L}_{\text{hard}},\qquad \alpha\in[0,1]. \label{eq:kd_total}
\end{align}
Here, $\alpha$ controls the trade-off between fitting labels (\emph{hard} supervision) and imitating the teacher (\emph{soft} supervision), while $T$ determines how much dark knowledge is revealed by the teacher distribution. \cite{Hinton2015Distilling,Gou2021KDSurvey}
The factor $T^{2}$ is conventionally used to offset the temperature-dependent scaling of gradients, helping keep the relative contribution of the soft term comparable across different $T$ values. \cite{Hinton2015Distilling,Gou2021KDSurvey}

% =========================
% Variational Quantum Circuits (VQCs)
% =========================
\subsection{Variational Quantum Circuits}\label{sec:vqc}

Variational quantum circuits (VQCs) are parameterized quantum models in which a quantum state is prepared by a sequence of gates that depend on trainable parameters, and the parameters are optimized by a classical routine to minimize a task-dependent loss. \cite{Cerezo2021VQA,Preskill2018NISQ}
This hybrid quantum--classical loop makes VQCs a natural candidate for learning in the noisy intermediate-scale quantum (NISQ) regime, where circuit depth and noise levels constrain fully quantum training. \cite{Preskill2018NISQ}

A typical VQC can be written as
\begin{equation}
\ket{\psi(\boldsymbol{\theta};\mathbf{x})}=U(\boldsymbol{\theta})\,U_{\mathrm{enc}}(\mathbf{x})\,\ket{0}^{\otimes n},
\end{equation}
where $U_{\mathrm{enc}}(\mathbf{x})$ encodes an input $\mathbf{x}$ into a quantum state and $U(\boldsymbol{\theta})$ is a trainable ansatz composed of parameterized single-qubit rotations and entangling gates. \cite{Schuld2018CircuitCentric,Havlicek2019FeatureSpaces,Cerezo2021VQA}
Model outputs are obtained by measuring expectation values of observables $\{O_m\}$,
\begin{equation}
s_m(\boldsymbol{\theta};\mathbf{x})=\bra{\psi(\boldsymbol{\theta};\mathbf{x})}O_m\ket{\psi(\boldsymbol{\theta};\mathbf{x})},
\end{equation}
and mapping them to class probabilities (e.g., via a sigmoid for binary classification or a softmax for multi-class settings). \cite{Schuld2018CircuitCentric}

In machine learning applications, VQCs can be interpreted as trainable feature maps into a high-dimensional Hilbert space, where measurement statistics define nonlinear decision functions. \cite{Havlicek2019FeatureSpaces,Franco2025QuantumPhases}
Their practical appeal is that they can be very parameter-efficient and hardware-aligned, but their trainability is affected by circuit design choices, noise, and optimization pathologies such as barren plateaus, motivating careful ansatz selection and depth control. \cite{Cerezo2021VQA}

In the context of model compression and distillation, a VQC can serve as a compact student that learns to approximate the predictive behavior of a large classical teacher while operating under explicit quantum resource constraints (number of qubits, circuit depth, and measurement budget). \cite{Hinton2015Distilling,Preskill2018NISQ}
This makes VQC students particularly relevant when deployment targets include quantum co-processors or when one seeks to study accuracy--efficiency trade-offs across heterogeneous learning paradigms.

% =========================
% Methodology
% =========================
\section{Methodology}\label{sec:method}

\begin{figure}[H]
\centering
\includegraphics[width=\textwidth]{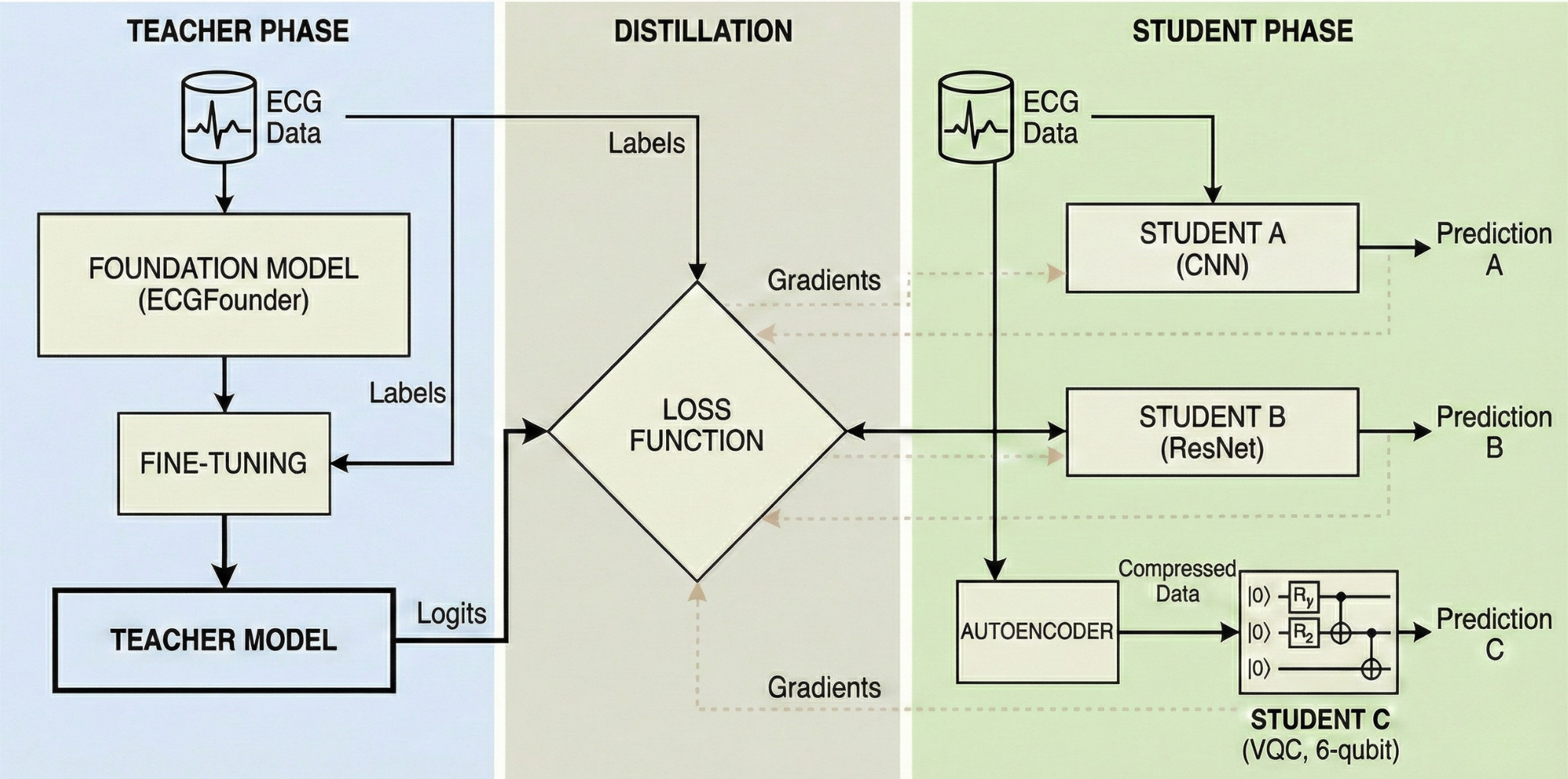}
\caption{Overview of the proposed pipeline. Left: a foundation model (ECGFounder) is fine-tuned to obtain a task-specific teacher. Center: a distillation loss combines hard labels and teacher logits to produce gradients. Right: three students (CNN, ResNet, and a compact VQC-based student) are trained under the same supervision signal for a controlled accuracy--efficiency comparison.}
\label{fig:distillation_pipeline}
\end{figure}

We compare distilled quantum and classical students under a single, standardized pipeline (Fig.~\ref{fig:distillation_pipeline}). The procedure comprises: (i) dataset preparation for binary ECG classification, (ii) ECGFounder fine-tuning to obtain a high-capacity teacher, (iii) knowledge distillation into three student families (CNN, compact ResNet, and a VQC-based student), and (iv) hyperparameter selection and evaluation with a fixed protocol. \cite{Li2024ECGFounder,Wagner2020PTBXL,Moody2001MITBIH,Hinton2015Distilling}

\FloatBarrier

\subsection{Data and Task Definition}\label{subsec:method_data}
We use PTB-XL and the MIT-BIH Arrhythmia Database and follow standard usage to define a binary ECG classification task. \cite{Wagner2020PTBXL,Moody2001MITBIH}
To enforce a consistent input setting across datasets and student architectures, we restrict the signals to a single channel (one lead) and standardize all recordings into fixed-format model inputs.

As part of preprocessing, we apply wavelet-based denoising to mitigate baseline wander and high-frequency artifacts prior to input construction. \cite{Addison2005WaveletECGReview,Zhang2005WaveletBaselineWander}
Specifically, each recording is decomposed via a discrete wavelet transform (DWT) and the detail coefficients are soft-thresholded using Donoho--Johnstone shrinkage principles (universal/SURE-type thresholding), with the noise level estimated from the median absolute deviation (MAD). \cite{Donoho1995SoftThresholding,DonohoJohnstone1995SureShrink,Poornachandra2008SubbandThresholdECG}

Model selection and evaluation are performed with stratified 5-fold cross-validation. Dataset splits are defined to prevent leakage (e.g., subject-level separation when applicable), and metrics are reported as averages across folds. \cite{Wagner2020PTBXL,Moody2001MITBIH}

\subsection{Teacher Fine-Tuning}\label{subsec:method_teacher}
ECGFounder is fine-tuned on each dataset to obtain a task-specialized teacher for \emph{binary} ECG classification. \cite{Li2024ECGFounder}
Starting from the pretrained ECGFounder encoder, we attach a binary classification head and optimize the model end-to-end with supervised cross-entropy (binary logistic loss) on the training splits. \cite{Li2024ECGFounder}
Model selection is performed using validation performance within the evaluation protocol (Section~\ref{subsec:method_eval}), and we retain the checkpoint that maximizes the chosen validation metric.
The resulting fine-tuned teacher produces sample-wise logits that are stored and subsequently used as soft targets for knowledge distillation of all student architectures. \cite{Hinton2015Distilling}

\subsection{Classical Students}\label{subsec:method_students}

We distill the teacher into two classical convolutional students: a compact ResNet-1D and a lightweight CNN-1D, both taking a single-lead ECG input $(1\times L)$ and producing a single binary logit. ResNet and CNN-based students are widely adopted in knowledge distillation as accuracy--efficiency baselines under constrained capacity, enabling controlled comparisons of how architectural inductive biases and parameter budgets affect the transfer of teacher behavior. \cite{Hinton2015Distilling,Romero2015FitNets,Zagoruyko2017AttentionTransfer,Yim2017FSP,Park2019RelationalKD,Tian2019CRD,Mirzadeh2020TeacherAssistant}

The ResNet-1D student follows the residual learning principle adapted to 1D temporal signals: a convolutional stem performs early feature extraction, followed by a small number of residual stages built from BasicBlocks (skip connections with $k{=}3$ Conv1D layers), with temporal downsampling introduced via strided convolutions at stage transitions. \cite{He2016ResNet}
The network ends with global average pooling and a linear layer that outputs the final logit. In contrast, the CNN-1D student is a lightweight feed-forward backbone composed of a short stack of strided Conv1D blocks (Conv1D + BN + ReLU) that progressively increases channel capacity while reducing temporal resolution, followed by global average pooling and a small fully connected head with dropout for regularization to produce the binary logit. \cite{Hinton2015Distilling,Gou2021KDSurvey}
Together, these two students provide complementary baselines: a residual architecture that improves optimization and feature reuse under moderate compression, and an ultra-compact convolutional model that probes the limits of distillation under a stringent parameter budget.

% =========================
% Quantum Student (Autoencoder + VQC)
% =========================
\subsection{Quantum Student}\label{subsec:method_quantum_student}

The quantum student couples a convolutional autoencoder with a 6-qubit variational quantum circuit (VQC) in order to interface high-dimensional ECG segments with a small qubit register. Given an ECG window of length $256$ samples, the encoder compresses the input into a latent vector of dimension $6$, which is then used as the VQC input. The encoder follows a lightweight 1D convolutional design (three Conv1D blocks with 16/32/64 filters, kernel size $5$), followed by flattening and a dense layer that outputs the 6-dimensional latent representation; the decoder mirrors this pathway to reconstruct the original window, encouraging the latent space to preserve morphology relevant for downstream classification while enforcing a strict dimensionality compatible with the quantum layer. \cite{Hinton2006Autoencoder}

\begin{figure}[!t]
\centering
\large
\setlength{\tabcolsep}{0pt}
\begin{quantikz}[row sep=0.33cm, column sep=0.70cm]
\lstick{$q_0$} &
\gate[wires=6,style=qboxgreen]{\textsf{ZZFeatureMap}} &
\gate[wires=6,style=qboxblue]{\textsf{EfficientSU2}} &
\meter{} \\
\lstick{$q_1$} & \qw & \qw & \meter{} \\
\lstick{$q_2$} & \qw & \qw & \meter{} \\
\lstick{$q_3$} & \qw & \qw & \meter{} \\
\lstick{$q_4$} & \qw & \qw & \meter{} \\
\lstick{$q_5$} & \qw & \qw & \meter{} \\
\end{quantikz}
\caption{Quantum-student circuit used in this work. A 6-qubit \textsf{ZZFeatureMap} (green) is followed by a hardware-efficient \textsf{EfficientSU2} ansatz (blue). All qubits are measured to form the output vector used for binary prediction/distillation.}
\label{fig:vqc_circuit}
\end{figure}
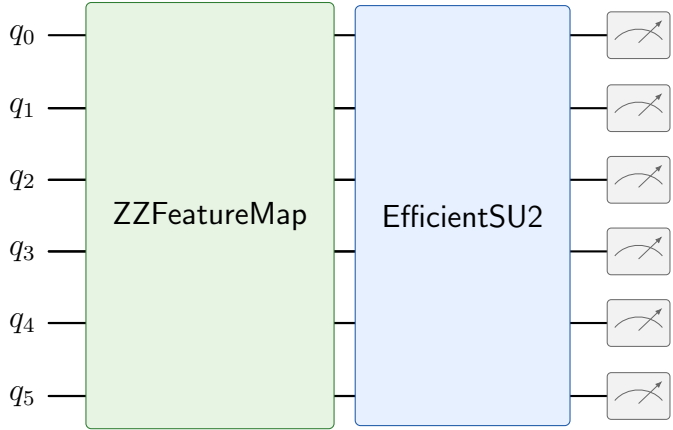

The VQC operates on $n=6$ qubits and is instantiated to match the Qiskit construction shown in Fig.~\ref{fig:vqc_circuit}. The latent vector $\mathbf{x}\in\mathbb{R}^{6}$ is embedded via a \textsf{ZZFeatureMap}, which implements a data re-uploading feature map combining input-dependent single-qubit phase rotations with pairwise $ZZ$ interactions under a linear entanglement pattern, yielding a nonlinear feature space for the subsequent variational circuit. \cite{Havlicek2019FeatureSpaces,Schuld2018CircuitCentric}
This is followed by a hardware-efficient \textsf{EfficientSU2} ansatz with trainable $R_y$ and $R_z$ rotations and linear entanglement, repeated for a small number of layers to keep depth compatible with NISQ constraints while retaining expressive capacity. \cite{Kandala2017HardwareEfficient,Cerezo2021VQA,Preskill2018NISQ}
Readout is performed by measuring all qubits (e.g., $\langle Z_i\rangle$ for $i=1,\dots,6$), producing a 6-dimensional output vector that is mapped to a binary prediction and also serves as the student output for distillation. \cite{Schuld2018CircuitCentric}
All quantum circuits are implemented in Qiskit and evaluated using a simulated backend to ensure controlled experiments under fixed circuit specifications and measurement settings. \cite{Qiskit2019}

\FloatBarrier

\subsection{Hyperparameter Search and Evaluation}\label{subsec:method_eval}
We perform a small grid search over the main distillation hyperparameters---the temperature $T$ and the mixing coefficient $\alpha$---and select the best configuration on the validation set for each student and dataset (Fig.~\ref{fig:kd_temperature_alpha}). The temperature controls the entropy of the teacher distribution used for soft supervision: larger $T$ softens the targets and makes relative logit structure more informative, which can stabilize learning for compact students and improve calibration when the teacher is highly confident. \cite{Hinton2015Distilling,Menon2021StatisticalDistillation,Huang2022StrongerTeacher,Li2023CTKD}
We evaluate $T\in\{2,4\}$, spanning a moderately softened regime that is commonly used in practical KD pipelines and that often yields measurable gains without overly flattening the supervisory signal. \cite{Hinton2015Distilling}

The mixing coefficient $\alpha$ controls the balance between matching the teacher outputs (soft supervision) and fitting ground-truth labels (hard supervision), thereby tuning how strongly the student relies on teacher-provided class relations versus direct label information. \cite{Hinton2015Distilling,Menon2021StatisticalDistillation,Gou2021KDSurvey}
For each temperature, we test $\alpha\in\{0.3,0.5,0.7\}$, yielding six $(\alpha,T)$ configurations per student. For each configuration, classical students are trained with Adam via standard backpropagation, while the quantum student is trained in a hybrid quantum--classical loop using SPSA, which is widely adopted for variational circuits due to its robustness in stochastic settings and low measurement overhead. \cite{Kingma2015Adam,Spall1998SPSA,Cerezo2021VQA}
Model selection is performed within the stratified 5-fold cross-validation protocol using validation performance, and final metrics are reported as averages over folds under the same evaluation procedure across all student families, enabling a controlled comparison of accuracy--efficiency trade-offs. \cite{Wagner2020PTBXL,Moody2001MITBIH}

% =========================
% Results
% =========================
\section{Results}\label{sec:results}

\subsection{Model complexity}\label{subsec:results_complexity}
\begin{figure}[!htbp]
\centering
\includegraphics[width=\textwidth]{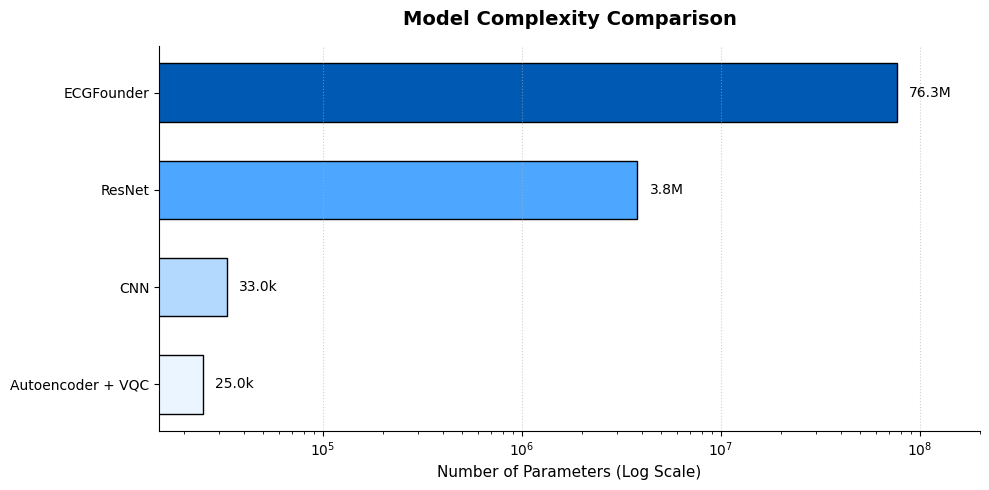}
\caption{Model complexity comparison (log scale). ECGFounder is the high-capacity teacher, while ResNet, CNN, and Autoencoder+VQC are distilled students with substantially reduced parameter budgets.}
\label{fig:model_complexity}
\end{figure}

Figure~\ref{fig:model_complexity} contextualizes the empirical results from a deployability perspective by comparing trainable parameter budgets (log scale). ECGFounder operates in a high-capacity regime ($76.3$M parameters), whereas the students are substantially smaller: ResNet-1D has $3.8$M parameters, and both CNN-1D and the Autoencoder+VQC pipeline fall in the $10^{4}$ range ($33.0$k and $25.0$k, respectively). This setting therefore shows a strong compression regime, ranging from an $\sim$6$\times$ reduction (ResNet) to nearly three orders of magnitude (CNN-1D and Autoencoder+VQC), and shifts attention to how effectively distillation preserves the teacher’s decision structure under constrained capacity.

These budgets correspond to distinct operational profiles. The teacher is well-suited to offline analysis or clinical backends where memory and latency constraints are less restrictive, but can be impractical for edge deployment. ResNet-1D represents an intermediate compromise, retaining sufficient representational capacity to approximate the teacher while enabling substantially cheaper inference, consistent with typical distillation objectives. In contrast, CNN-1D and Autoencoder+VQC target an aggressively lightweight regime, where preserving subtle morphological cues under limited capacity becomes the primary challenge; in this regime, the distillation signal is particularly important for stabilizing the learned decision rule and its calibration.

For the Autoencoder+VQC student, the parameter count is dominated by the convolutional autoencoder, which has approximately $25$k trainable parameters, whereas the VQC contributes only $36$ trainable parameters from the EfficientSU2 ansatz. Consequently, the total budget reported for the pipeline is largely determined by the encoder--decoder weights, with only a marginal contribution from the variational circuit parameters.

\subsection{Quantitative results}\label{subsec:results_tables}
Tables~\ref{tab:mit_results} and~\ref{tab:ptb_results} report mean test-set metrics (accuracy, precision, recall, and F1 score) for each student at $T=2$ and $T=4$, together with the fine-tuned teacher (ECGFounder) as a reference. Across both datasets, the teacher sets a clear upper bound, while KD produces compact students that retain a substantial fraction of the teacher’s F1 under large parameter reductions. A consistent trend is that students preserve high recall more readily than precision, indicating that the dominant degradation mode under compression is increased false positives rather than missed detections.

\subsubsection{Results on MIT-BIH Arrhythmia}
On MIT-BIH (Table~\ref{tab:mit_results}), the teacher achieves very high recall ($0.9900$) and the best overall balance with F1 Score ($0.8400$). Distilled students remain in a high-recall regime (recall $\sim$0.93--0.98), indicating that knowledge distillation preserves much of the teacher’s sensitivity on this dataset. However, all students exhibit a noticeable precision reduction relative to the teacher, which contributes to the lower accuracies and highlights MIT-BIH as the more challenging setting for compressed learners, where the remaining gap to the teacher is reflected primarily in precision and F1 Score.

Within this regime, differences between student families are modest. At $T=2$, the Autoencoder+VQC pipeline attains the best student accuracy and F1 (accuracy $0.6879$, F1 $0.8078$), while ResNet achieves the highest recall ($0.9763$). At $T=4$, performance becomes tightly clustered across students, with near-identical F1 scores ($\sim$0.795$-$0.797), suggesting that under a reasonable distillation configuration, student choice has a smaller impact than the dataset difficulty itself.

Importantly, the Autoencoder+VQC pipeline remains competitive with the classical students despite employing an extremely compact variational circuit: the VQC includes only $36$ trainable parameters. While most trainable capacity in the pipeline resides in the convolutional autoencoder, the quantum layer acts as a highly parameter-efficient decision module and still supports student-level performance comparable to ResNet-1D and CNN-1D on MIT-BIH, particularly in terms of accuracy and F1 Score at $T=2$.

\begin{table}[!htbp]
\centering
\caption{Mean results on MIT-BIH Arrhythmia for different temperatures ($T$).}
\label{tab:mit_results}
\begin{tabular}{llcccc}
\toprule
\textbf{Temp} & \textbf{Model} & \textbf{Accuracy} & \textbf{Precision} & \textbf{Recall} & \textbf{F1 Score} \\
\midrule
Teacher
 & ECGFounder & 0.8000 & 0.7900 & 0.9900 & 0.8400 \\
\midrule
$T=2$
 & ResNet & 0.6757 & 0.6775 & \textbf{0.9763} & 0.8033 \\
 & VQC    & \textbf{0.6879} & \textbf{0.7183} & 0.9392 & \textbf{0.8078} \\
 & CNN    & 0.6693 & 0.6757 & 0.9502 & 0.7971 \\
\midrule
$T=4$
 & ResNet & 0.6796 & 0.6848 & 0.9397 & 0.7952 \\
 & VQC    & \textbf{0.6817} & \textbf{0.6950} & 0.9333 & 0.7965 \\
 & CNN    & 0.6693 & 0.6864 & \textbf{0.9502} & \textbf{0.7971} \\
\bottomrule
\end{tabular}
\end{table}

% NÃO use \FloatBarrier aqui

\subsubsection{Results on PTB-XL}
On PTB-XL (Table~\ref{tab:ptb_results}), the teacher again provides the strongest reference performance (precision $0.8626$, recall $0.9512$, F1 $0.8179$). Relative to MIT-BIH, students typically achieve higher absolute accuracies on PTB-XL and a smaller accuracy gap to the teacher under the same protocol. In contrast, student recall is \emph{closer} to the teacher on MIT-BIH than on PTB-XL, indicating that the two datasets differ in how compression impacts sensitivity versus calibration.

In this regime, capacity differences are reflected more clearly among the classical baselines: ResNet-1D emerges as the strongest classical student overall, whereas CNN-1D trails, consistent with the expected accuracy--capacity trade-off for the smallest model.

The comparison between ResNet-1D and Autoencoder+VQC reveals a temperature-dependent precision--recall shift. At $T=2$, ResNet-1D leads accuracy and F1 (accuracy $0.7480$, F1 $0.7897$), while Autoencoder+VQC achieves the highest recall ($0.8310$), corresponding to a more sensitive operating point. At $T=4$, Autoencoder+VQC improves accuracy and precision (accuracy $0.7503$, precision $0.7766$) but reduces recall ($0.7975$), indicating a shift toward a more conservative decision regime. ResNet-1D remains marginally best in recall and slightly best in F1, while CNN-1D stays below both across temperatures.

Finally, the VQC-based student maintains competitive performance on PTB-XL despite operating with an extremely small number of trainable parameters in the variational circuit.

\begin{table}[!htbp]
\centering
\caption{Mean results on PTB-XL for different temperatures ($T$).}
\label{tab:ptb_results}
\begin{tabular}{llcccc}
\toprule
\textbf{Temp} & \textbf{Model} & \textbf{Accuracy} & \textbf{Precision} & \textbf{Recall} & \textbf{F1 Score} \\
\midrule
Teacher
 & ECGFounder & 0.8063 & 0.8626 & 0.9512 & 0.8179 \\
\midrule
$T=2$
 & ResNet & \textbf{0.7480} & \textbf{0.7607} & 0.8223 & \textbf{0.7897} \\
 & VQC    & 0.7293 & 0.7366 & \textbf{0.8310} & 0.7798 \\
 & CNN    & 0.7207 & 0.7367 & 0.8032 & 0.7680 \\
\midrule
$T=4$
 & ResNet & 0.7417 & 0.7486 & \textbf{0.8316} & \textbf{0.7876} \\
 & VQC    & \textbf{0.7503} & \textbf{0.7766} & 0.7975 & 0.7861 \\
 & CNN    & 0.7123 & 0.7268 & 0.8027 & 0.7626 \\
\bottomrule
\end{tabular}
\end{table}

\subsection{Distillation hyperparameters: temperature and $\alpha$}\label{subsec:results_hparams}
\begin{figure}[!htbp]
\centering
\includegraphics[width=\textwidth]{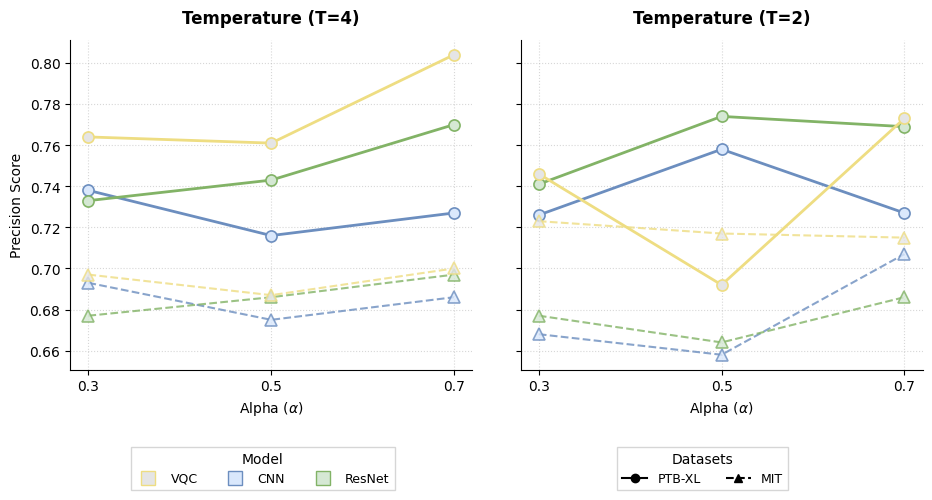}
\caption{Precision score as a function of the distillation hyperparameters $\alpha$ and temperature $T$ for PTB-XL and MIT-BIH, comparing VQC, CNN, and ResNet students. Solid lines denote PTB-XL and dashed lines denote MIT-BIH.}
\label{fig:kd_temperature_alpha}
\end{figure}

Figure~\ref{fig:kd_temperature_alpha} complements the table-level outcomes by isolating how precision varies with the KD hyperparameters. Precision is reported as a function of $\alpha\in\{0.3,0.5,0.7\}$ for $T\in\{2,4\}$, with PTB-XL in solid lines and MIT-BIH in dashed lines. In line with Tables~\ref{tab:mit_results}--\ref{tab:ptb_results}, PTB-XL tends to yield higher precision than MIT-BIH across students and settings, which may indicate that MIT-BIH is a more demanding regime for calibration under compression.

The temperature effect appears to be most pronounced for the Autoencoder+VQC student. For $T=4$, the VQC-based student may remain the top-precision model relative to the other students across the tested $\alpha$ values, with precision tending to improve as $\alpha$ increases and peaking at $\alpha=0.7$ on PTB-XL. A similar advantage may persist at $T=2$, except for $\alpha=0.5$ on PTB-XL, where the VQC student shows a noticeable descent followed by recovery at $\alpha=0.7$. This pattern may indicate that higher temperature provides more informative soft targets for the most constrained student, while the choice of $\alpha$ can influence training stability by anchoring learning more strongly to hard-label supervision.

In contrast, ResNet-1D and CNN-1D appear to exhibit smoother trends across $\alpha$ and smaller differences between $T=2$ and $T=4$, consistent with the more stable behavior suggested by the quantitative results in Tables~\ref{tab:mit_results}--\ref{tab:ptb_results}.

\section{Conclusion}\label{sec:conclusion}

In this work, we investigated whether a large ECG foundation model can be effectively compressed into highly compact students for binary ECG classification on PTB-XL and MIT-BIH. Fine-tuning ECGFounder provided a strong teacher baseline, while knowledge distillation enabled substantial reductions in trainable parameters---from $76.3$M parameters down to $3.8$M (ResNet-1D), $33.0$k (CNN-1D), and $25.0$k (Autoencoder+VQC)---while retaining competitive performance relative to the teacher.

Across both datasets, the distilled students preserved a high-recall operating regime, with performance depending on the distillation hyperparameters. Temperature and mixing coefficient effects were more evident on precision, and the $T=4$ setting tended to yield more stable precision gains. Within this setting, the Autoencoder+VQC pipeline remained consistently competitive with the classical students, despite using an extremely compact variational circuit with only $36$ trainable parameters, which may indicate that the quantum layer can serve as a parameter-efficient decision module within the distillation pipeline.

Overall, these results suggest that distillation can transfer a substantial fraction of a foundation teacher's behavior not only into compact classical models, but also into a quantum-ready student that couples a low-dimensional convolutional autoencoder with a shallow variational circuit. The competitive performance observed for the VQC-based student supports the view that quantum students can be viable alternatives under strong compression, motivating further studies on hybrid architectures, robustness under dataset shift, and execution on noisy quantum hardware to quantify the gap between simulation and NISQ devices.

\subsection*{Funding}
This study was funded by the Fundação de Amparo à Pesquisa do Estado de São Paulo (FAPESP) through Project Nos. 2025/19585-6 (G.S.F.), 2023/04987-6 and 2024/00998-6 (F.F.), and by the Conselho Nacional de Desenvolvimento Científico e Tecnológico (CNPq) through Project No. 408884/2024-0 (F.F.).
%==============
% References
% =========================
\bibliographystyle{bst/sn-aps}
\bibliography{sn-bibliography}

\end{document}